\documentclass[12pt,preprint]{aastex}

\usepackage{lscape}

\newcommand{\etal}{{et al.}}

\begin{document}

\newcommand{\gsim}{\raisebox{-.4ex}{$\stackrel{>}{\scriptstyle \sim}$}}
\newcommand{\lsim}{\raisebox{-.4ex}{$\stackrel{<}{\scriptstyle \sim}$}}
\newcommand{\psim}{\raisebox{-.4ex}{$\stackrel{\propto}{\scriptstyle \sim}$}}
\newcommand{\kms}{\mbox{km~s$^{-1}$}}
\newcommand{\s}{\mbox{$''$}}
\newcommand{\mloss}{\mbox{$\dot{M}$}}
\newcommand{\my}{\mbox{$M_{\odot}$~yr$^{-1}$}}
\newcommand{\ls}{\mbox{$L_{\odot}$}}
\newcommand{\ms}{\mbox{$M_{\odot}$}}
\newcommand{\mm}{\mbox{$\mu$m}}
\newcommand{\water}{\mbox{H$_2$O}}
\newcommand{\methanol}{\mbox{CH$_3$OH}}
\def\arcdeg{\hbox{$^\circ$}}
\newcommand{\secp}{\mbox{\rlap{.}$''$}}%

\shortauthors{Dodson, et al.}
\shorttitle{Astrometric observations of H$_2$O and SiO masers in
  R\,LMi using KVNopen }

\title{Astrometrically Registered Simultaneous Observations of the 22 GHz \water\  and the 43GHz 
  SiO masers towards R Leonis Minoris using KVN and 
Source/Frequency Phase Referencing} 

 \author{Richard \textsc{Dodson}\altaffilmark{1,2},
   Mar\'{\i}a J. \textsc{Rioja}\altaffilmark{1,2,3},
   Tae-Hyun \textsc{Jung}\altaffilmark{1},
Bong-Won Sohn\altaffilmark{1},
Do-Young Byun\altaffilmark{1}, 
Se-Hyung Cho\altaffilmark{1},
Sang-Sung Lee\altaffilmark{1},
Jongsoo Kim\altaffilmark{1},
Kee-Tae Kim\altaffilmark{1},
Chung-Sik Oh\altaffilmark{1},
Seog-Tae Han\altaffilmark{1},
Do-Heung Je\altaffilmark{1},
Moon-Hee Chung\altaffilmark{1},
Seog-Oh Wi\altaffilmark{1},
Jiman Kang\altaffilmark{1},
Jung-Won Lee\altaffilmark{1},
Hyunsoo Chung\altaffilmark{1},
Hyo-Ryoung Kim\altaffilmark{1},
Hyun-Goo Kim\altaffilmark{1},
Chang-Hoon Lee\altaffilmark{1},
Duk-Gyoo Roh\altaffilmark{1},
Se-Jin Oh\altaffilmark{1},
Jae-Hwan Yeom\altaffilmark{1},
Min-Gyu Song\altaffilmark{1},
Yong-Woo Kang\altaffilmark{1}
}
 \affil{$^1$ Korea Astronomy and Space Science Institute, Daedeokdae-ro 776, Yuseong-gu, Daejeon 305-348, Korea}
 \affil{$^2$ International Centre for Radio Astronomy Research, M468,
The University of Western Australia, 35 Stirling Hwy, Crawley, Western Australia, 6009}
 \affil{$^3$ Observatorio Astron\'omico Nacional (IGN), Alfonso XII, 3 y 5, 28014 Madrid, Spain} 
 \email{rdodson@kasi.re.kr}



\begin{abstract}

  Oxygen-rich Asymptotic Giant Branch (AGB) stars can be intense emitters
  of SiO ($v$=1 and
  2, J=1$\rightarrow$0) and H$_2$O maser lines at 43 and 22 GHz, respectively. VLBI observations of the maser emission
  provide a unique tool to probe the innermost layers of the
  circumstellar envelopes in AGB stars.  Nevertheless, the
  difficulties in achieving astrometrically aligned \water\ and $v$=1 and
  $v$=2 SiO maser maps have traditionally limited the physical
  constraints that can be placed on
  the SiO maser  pumping mechanism.
  We present phase referenced simultaneous spectral-line VLBI images
  for the SiO $v$=1 and $v$=2,  J=1$\rightarrow$0, and H$_2$O maser emission around
  the AGB star R\,LMi,
  obtained from the Korean VLBI Network (KVN). The simultaneous
  multi-channel receivers of the KVN offer
  great possibilities for astrometry in the frequency domain. With this
  facility we have produced images with bona-fide absolute astrometric
  registration between high frequency maser transitions of different
  species to provide the positions of the \water\ maser emission, and the
  centre of the SiO maser emission, and hence reducing the uncertainty
  in the proper motion for R\,LMi by an order of magnitude over that
  from Hipparcos.  This is the first successful demonstration of source frequency phase
  referencing for mm-VLBI spectral-line observations and also where the ratio between the
  frequencies is not an integer. 

\end{abstract}

\section{Introduction}

O-rich Asymptotic Giant Branch (AGB) stars can be intense emitters of SiO
and H$_2$O maser lines. 
There are many molecular lines which mase in AGB stars
\citep{elitzur_92}, in particular those of SiO (for example
\cite{gray_99}) and \water\ (for example \cite{water_agb}).
The most commonly studied are the 43-GHz SiO lines ($v$=1 and 2,
$J$=1$\rightarrow$0) and 22-GHz \water ,
($J_{K_-,K_+}$=6$_{1,6}$--5$_{2,3}$). Simultaneous frequency single dish
monitoring of these lines, along with others, are an active research
effort at the KVN, e.g. \citet{sio_h2o_kasi}.
%
SiO masers have very high excitation levels and so appear at a
distance of a few stellar radii, $\sim$10$^{14}$ cm, resulting in a
more or less circular structure. H$_2$O emission is found further from
the star, $\sim$10$^{15}$ cm, with less well defined structures. A
combined study of both masers can therefore produce a very accurate
description of the structure and kinetics of these inner layers. In
the inner-most regions of circumstellar shells, from which the whole
circumstellar envelope will be formed, the dust grains are still
growing and the gas has not yet attained its final expansion velocity,
since expansion is supposed to be powered by radiation pressure onto
the grains. The dynamics in the SiO emitting region is dominated by
pulsation, which propagates from the photosphere via shocks, and by
the first stages of outwards acceleration \citep{bowen88,hump02,gray_09}.
The \water\ emission, on the other hand, occurs where the acceleration process is almost
completed.
Their combined study therefore provides a powerful tool for the understanding the mass loss and the
envelope formation processes in AGB stars.

To understand these mass loss processes we need to derive the underlying
physical conditions from the observations of the maser emission. The
theory for \water\ maser production is understood to be due to collisions with hydrogen
\citep{elitzur_water}, however the
theory of the SiO maser excitation is the weakest link in the
logical chain. For the SiO molecule there are currently two models,
one based on radiative pumping \citep{buj94} from the NIR emission
of the central star, the other on collisional pumping
\citep{elitzur_80,hum2000} 
via the shock wave driven by stellar pulsation. If the latter is the
correct model the prediction is that there would be no or little
separation between the  $v$=1 and $v$=2 SiO
emission regions, whereas for the former the prediction is that there
would be such a separation.  
Therefore the relative distribution of the spots at both the $v$=1 and $v$=2 SiO
transitions is important to discriminate between the different pumping mechanisms. Existing multi-transitional maps \citep{desmurs00, yi05, bob_05, soria04, soria05,
  soria07, rioja_08, cot_04, cot_06, cot_08, cot_09a, cot_09b,
  cot_10a, cot_10b, cot_11, richter13, choi_08, kamohara_10} suggest that the spots of these
lines are generated in close proximity. One school of thought, however, reports that
these emissions tend to be separated by about an AU (a mas at
typical distances) or that the typical radius for the SiO transitions
are different, thereby favouring the radiative pumping model; the other finds the
emission coincident, thereby favouring the collision pumping model. 
However in the majority of cases the basis for these conclusions is either
the assumption of a common origin for the centres of the SiO maser emissions or
from alignment performed on cross correlation of the
images. Neither method provides {\it bona-fide} astrometry. 
Only \citet{rioja_08, choi_08} and \citet{kamohara_10} provide robust bona-fide
astrometric registration between the SiO transitions, all from VERA
observations. In \citet{rioja_08} we observed R\,LMi and found only
one site out of 17 spots produced overlapping $v$=1 and $v$=2 emission.
\citet{kamohara_10} and \citet{choi_08} astrometrically observed
R\,Aquilae and VY\,CMa respectively, also with VERA observations. Whilst
\citet{choi_08} had astrometrical information no comparison
between the transitions was made. \citet{kamohara_10} showed that only a few
percent of the spots showed overlap between the $v$=1 and $v$=2 emission.
Of the other observations only \citet{bob_05} and \citet{yi05}
attempted astrometry but these had narrow
limited bandwidth coverage limiting the delay (and thus positional)
accuracy or had a poor absolute coordinate position limiting the final
relative positional accuracy.
%
%

In general the stellar positions are quite accurately
measured for many AGB stars, with data from the Hipparcos satellite
\citep{perry_97} providing typical precision of 1\,mas and it is
expected that GAIA \citep{gaia} will provide precisions two orders of
magnitude higher. It is therefore the VLBI studies
of the SiO and H$_2$O maser spot distributions which are lacking
accurate astrometric information in their relative and absolute
coordinates of the maps of the different lines. Given the difficulties
in providing these measurements with the existing facilities we have
been investigating new astrometric methods.

The innovative multi-channel receiver \citep{han_08} of the Korean VLBI Network
(KVN) \citep{sslee_kvn} is designed to allow the transfer of the calibration solutions
derived from the measurements on one
frequency band (or `channel' in their nomenclature) to the data from another frequency band. 
This provides the ideal design for experiments which benefit from
simultaneous observations at different frequencies or spectral
transitions. 
The KVN backend, combined with the Source Frequency Phase Referencing
(SFPR) technique \citep{vlba_31,rioja_11a,rioja_11b}, allows astrometrical observations even at the
highest frequencies. We know of no demonstrated upper limit and it
would be expected to work as long as the tropospheric contributions
were non-dispersive. Tests are on-going with ALMA \citep{alma_test} at
frequencies as high as 350 GHz, and so far have been successful.
In \citet{rioja_14} we make a detailed comparison between the KVN and
the VLBA for continuum sources. The frequencies
of spectral line emission do
not necessarily fall on integer ratios and in this paper we focus only
on the
additional considerations and steps required for the analytical method in this case. Future
publications with multiple epochs of data will be more suitable to
explore the physical interpretation of the observations.

We present here the results from simultaneous KVN observations of
SiO and H$_2$O maser emission in R\,LMi,
and the first ever bona-fide absolute astrometric alignement between
\water\  and  ($v$=1 and $v$=2, J=1$\rightarrow$0) SiO
maser emission derived from the SFPR  astrometric analysis. 
R\,LMi is an O-rich Mira-type variable, with a pulsation period of
about 
373 days \citep{pardo_04} and a spectral type ranging between M6.5, at the optical
maximum, and M9.0, at the minimum \citep{keenan_74}. Its distance is assumed to be $\sim$
417 pc \citep{RLMi_distance}, derived from the well-known
period-luminosity relation \citep{PL-rel}. R\,LMi is
a well known emitter in \water\ and SiO lines. In particular it was
accurately monitored by \citet{pardo_04}, who found the periodic
variation in radio flux in-phase with the IR cycle, typical of Mira-type
stars. 

In our previous publications (e.g. \citet{rioja_11a}) 
we strongly suggested that SFPR required
an integer frequency ratio between the calibrating and the target
frequencies. This is not always possible, particularly for spectral
sources, where the line rest frequency sets the target
frequencies.
This effectively limited the spectral line science targets to be SiO masers, as the
rotational emission modes for this molecule have almost exact integer
frequency ratios. 
Here however we successfully phase
referenced the SiO maser emission with measurements on \water\  maser
emission, where the frequency ratio is around 1.9. This small
offset from integer had been found to be sufficient to prevent success
in previous attempts to achieve astrometry
\citep{dodson_eavw_11}. These early attempts failed because the
calibration chain for the astrometric observations were insufficiently
rigorous, as will be explained in the following sections.  
%

The contents of the paper are organised as follows.  The observational
setup is described in Sect.\ \ref{sec:obs}. In Sect.\,\ref{sec:red} we describe the special
data reduction strategy used to preserve the relative astrometry
between the maser transitions. In Sect.\,\ref{sec:res} we briefly present
the first astrometrically aligned maps and in Sect.\,\ref{sec:dis} we
discuss the implications of this new method for the field.

\section{Observations}
\label{sec:obs}

We carried out simultaneous dual-frequency observations in early test time with the KVN
(2011, March 4th) using all three antennas at Yonsei, Ulsan and
Tamna. We recorded 8 narrow bands (intermediate-frequency bands, or IFs)
of Left Hand Circular (LHC) 16MHz bandwidth from the 22GHz system
and the same from the 43GHz system. The IFs were set as widely separated
and as randomly distributed as possible across the available system bandwidth of
500MHz at each band, whilst including the target transitions, in order
to suppress the delay sidelobes. At 22GHz the IFs were with lower band edges at 21.810,
21.826, 21.858, 21.922, 22.050, 22.114, 22.162, 22.226\,GHz. This places the
\water\  emission close to the centre of IF\,8. For the 43GHz system we
had IFs with lower band edges at 42.810, 42.826, 42.858, 42.922, 43.050,
43.114, 43.162, 43.226\,GHz, with the $v$=2 SiO maser in IF\,1 and $v$=1 SiO
maser in IF\,6. This distribution ensured that the first delay
sidelobe, at $\sim$16\,$\mu$sec, was less than 70\% of the main
peak.

We alternated observations between the target R\,LMi and the bright continuum
fringe-finder and calibrator source, 4C39.25 (J0927+3902), with 2 minute
scans on each. The angular separation between these two sources is
5.9$^o$ on the sky. 
The correlation was done with the DiFX correlator \citep{difx} with 1
second averaging and a
spectral resolution of 1024 channels per IF,
yielding maximum velocity resolutions of 0.1\,\kms\ and 0.2\,\kms\ for the line
observations at 43 and 22 GHz, respectively, when using no spectral smoothing.

\section{Data reduction for Source Frequency Phase Referencing Analysis} 
\label{sec:red}
\subsection{Basis of the Source Frequency Phase Referencing method}

The full details of SFPR are
provided in two VLBA memos \citep{vlba_31,vlba_32} and in two
journal papers \citep{rioja_11a,rioja_11b}. 
The method is generally applicable, but has particular relevance in
cases where the frequency is so high that conventional phase
referencing (PR) can not, or struggles to, succeed.  In PR the
antennas nod rapidly between a calibrator (assumed to be achromatic
with no core-shift of its own, in both PR and SFPR) and the target
\citep{alef, vlba_pr}. For frequencies above
22GHz the coherence time becomes so short that it becomes increasingly
difficult to slew from one source to the other and back within this
period. Additionally the number of calibrators strong enough to be
detected, yet within the same tropospheric `patch' as the target, rapidly
approaches zero.
SFPR avoids this limit by observing at multiple frequencies and assuming that the atmosphere is dominated by
non-dispersive contributions.  At mm-wavelengths this is certainly the
case as the troposphere is the limiting source of error, and the
troposphere is non-dispersive. 
By self-calibrating on the observations of the target source at the low
(reference) frequency and applying those solutions to the higher
frequency data we correct for the dominant non-dispersive terms.
We call the resultant calibrated high frequency dataset ``Frequency Phase Transferred'' ({\it
  hereafter} FPT), or {\it troposphere-free}, target dataset.  However
the remaining dispersive errors prevent the recovery of the position of the
target source, using a Fourier inversion, at this stage.
The residual dispersive contributions arising from the ionospheric propagation
and other effects need to be removed using the alternating observations of the second
source. This source can be visited less frequently and can lie
significantly further from the target than in conventional PR, as the residual terms these
observations correct for are much weaker and vary much more slowly than the dominant
tropospheric and other non-dispersive effects. After this second
calibration step the 
resultant {\sc sfpr}-calibrated data-sets can be Fourier inverted
to provide the astrometrically registered target image between the two
observing frequencies. 

\begin{figure}
  \begin{center}
    \includegraphics[width=0.5\textwidth,angle=-90]{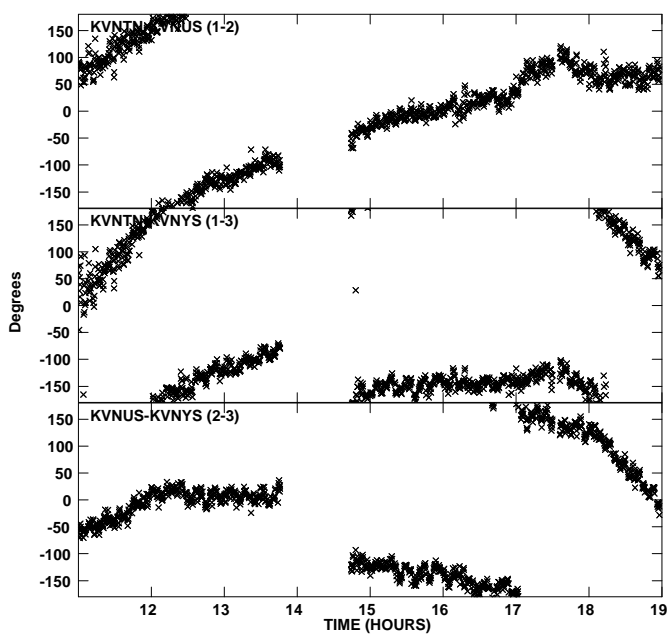}
  \end{center}
   \caption{Residual visibility phases of 4C39.25 at 43GHz after the Frequency
     Phase Transfer step, that is the application of the scaled corrections
     derived at 22GHz on the simultaneous low frequency
     observations. Whilst the phases are disciplined and have coherence
     timescales of the order of an hour, the remaining dispersive terms
     prevent direct imaging.} 
\label{fig:phase}
\end{figure}

As stated the first step does not
provide astrometry-ready data, as shown in Figure \ref{fig:phase}. Following equation 2 in \cite{rioja_11a},
the FPT dataset has residual phases $\phi^{{\rm FPT}}$, that can be
expressed as the sum of the contributions:

\begin{center}
\begin{eqnarray}
\phi^{{\rm FPT}} = 
\phi^{{\rm high}} - R\,.\, \phi_{{\rm self-cal}}^{{\rm low}} 
= \phi_{{\rm str}}^{{\rm high}} \nonumber \\ 
+ (\phi_{{\rm geo}}^{{\rm high}} - R\,.\, \phi_{{\rm geo}}^{{\rm low}}) 
+ (\phi_{{\rm tro}}^{{\rm high}} - R\,.\, \phi_{{\rm tro}}^{{\rm low}}) \nonumber \\
+ (\phi_{{\rm ion}}^{{\rm high}} - R\,.\, \phi_{{\rm ion}}^{{\rm low}})  
+ (\phi_{{\rm inst}}^{{\rm high}} - R\,.\, \phi_{{\rm inst}}^{{\rm low}}) \nonumber \\
+ 2\pi (n^{{\rm high}}-R\,.\,n^{{\rm low}})  ,
\end{eqnarray} 
\end{center}

where $\phi^{{\rm low}}_{{\rm geo}}, \phi^{{\rm low}}_{{\rm tro}},
\phi^{{\rm low}}_{{\rm ion}}$, $\phi^{{\rm low}}_{{\rm inst}}$, $\phi^{{\rm high}}_{{\rm geo}}, \phi^{{\rm high}}_{{\rm tro}},
\phi^{{\rm high}}_{{\rm ion}}$ and $\phi^{{\rm high}}_{{\rm inst}}$ are
the contributions arising from geometric, tropospheric, ionospheric
and instrumental inadequacies in the delay model for, respectively,
either the ${\rm low}$ or ${\rm high}$ frequency. $n^{\rm low}$ and $n^{\rm high}$ 
are the integer number of full phase rotations, or ambiguities in the
phase, for each of the frequencies and $R$ is the frequency ratio between the high and low
frequencies.
At mm-wavelengths the geometric and tropospheric phase residual terms
will scale with frequency and therefore these terms in parenthesis
will cancel; instrumental terms are assumed to have been dealt with
using observations of a primary calibrator. Whereas the ionospheric
terms, although non-zero, are slow-changing in time and equal over a
large angular separation.
When $R$ is an integer the phases from the ambiguity terms $n$ are
always integer multiples of 2$\pi$, so are not relevant to the analysis.
However when $R$ is not an integer phase jumps are introduced every
time $n^{{\rm low}}$ changes. That is at every introduction of an
ambiguity in phase there will be a step in the phase of $\phi^{{\rm FPT}}$,
which can not be corrected from the observations of the second source,
as for the calibrator the number of ambiguities is independent along
that different line of sight. Clearly the solution for non-integer
frequency ratio SFPR is equivalent to that required to avoid
introducing any untracked lobe rotations in the visibility data.
Below we describe our data reduction procedure to retain the
astrometric information, even when the ratio between the observing frequencies is not an
integer number. 

\subsection{Source Frequency Phase Referencing with non-integer
  frequency ratios}

\begin{figure}
  \begin{center}
    \includegraphics[width=0.5\textwidth]{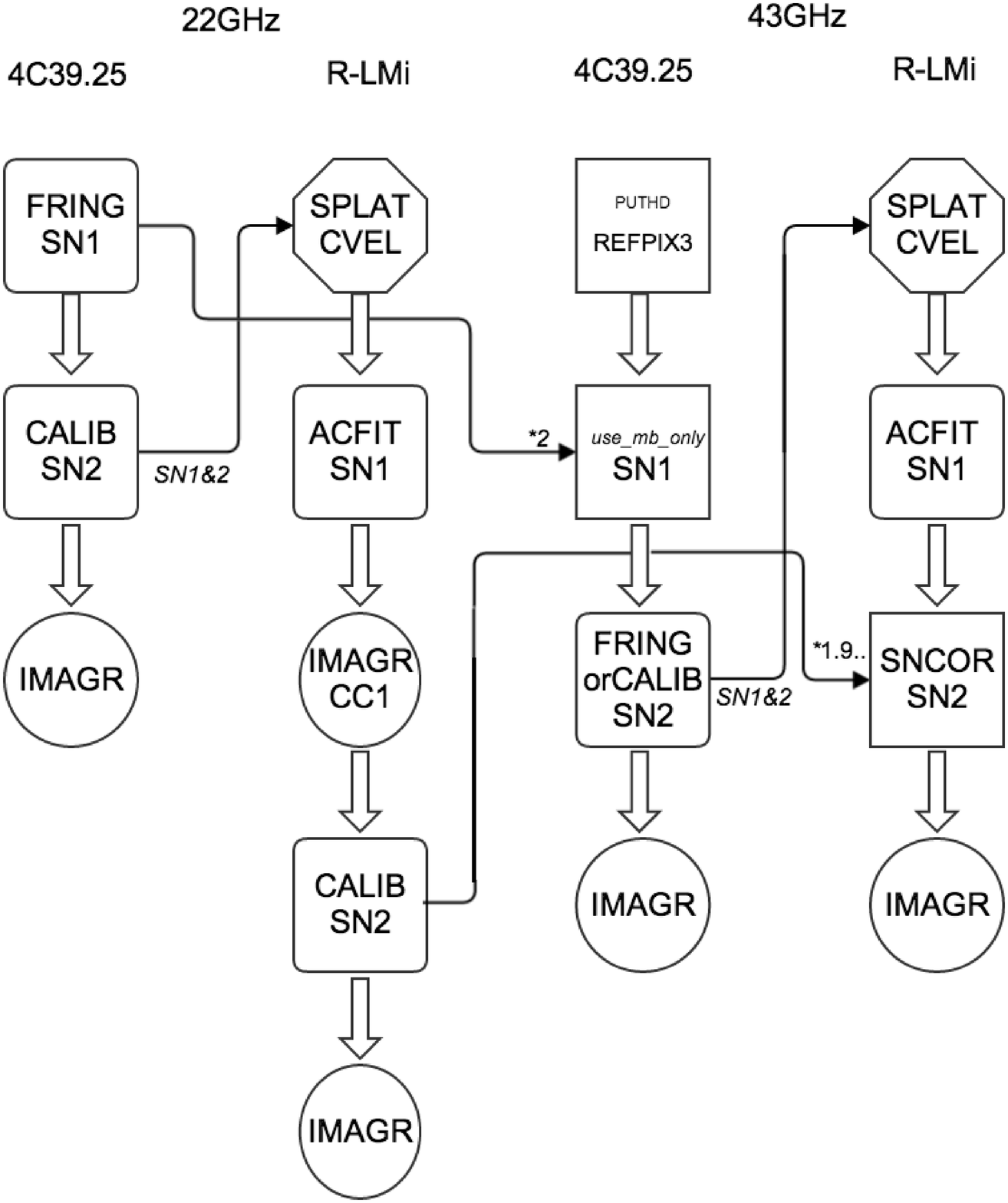}
  \end{center}
   \caption{Schematic flow for the non-integer SFPR data
     reduction. `Rounded squares' (e.g. FRING or CALIB) are calibration
   derivations, `Squares' are operations on those calibration
   values. `Hexagons' are operations on the {\em uv}-data and circles
   are the `inversion' of the {\em uv}-data to form the image. In all
   cases the AIPS task and the calibration table are specified.} 
\label{fig:flow}
\end{figure}

We used the NRAO AIPS software package for the data reduction \citep{aips}. The
information on measured system temperatures, telescope gains, and
estimated bandpass corrections for the individual antennae were used to
calibrate the raw correlation coefficients of the continuum
calibrators and of the target spectral line source. 
Figure \ref{fig:flow} shows a schematic of the steps after this
initial calibration for each source and for each frequency, indicating
the transfer of the calibration tables. Each of the four columns in that figure
represents a different stage in the analysis.

Column 1 is for the reference source at the reference frequency,
i.e. 4C39.25 at 22 GHz.  
The first step is to use the AIPS task FRING, which is a global self-calibration
algorithm \citep{cotton_fring_95}, to estimate residual antenna-based VLBI phases, and its
partial derivatives, with respect to frequency (group delay,
$\tau$), and time (phase delay rate), on the calibrator (4C39.25) data set. 
These terms result from unaccounted contributions from the atmospheric
propagation and from errors in the array geometry during the data
correlation. 
%
The performance of the digital filters in the KVN optics
\citep{kvn_optics} and backend \citep{kvn_backend} is
extremely good, particularly in the absence of instrumental phase 
offsets introduced by the electronics at each IF and band. This
enables the straight-forward use of all the IFs as an effective single
bandwidth for the estimation of a more precise group delay (since
$\sigma_{\tau}$ $\propto$ 1/bandwidth) \citep{cotton_schwab}. With a reasonable SNR and
432MHz spanned bandwidth we expect delay accuracies of tens of picoseconds. 
We used the delays derived from
this calibrator and frequency, solving for a single delay across all
the IFs, in all the subsequent analysis of the other sources and frequencies. 
Finally for this source and frequency we applied the small ($<$10$^o$) phase
corrections which were constant across the whole experiment, to align the IFs. These
were derived with the phase self-calibration task CALIB; normally this would be done with FRING on a
single scan of a strong calibrator but we applied this as a refinement
after the application of the initial calibration across all the
observations as the single band delay corrections required for the KVN
receiver were essentially zero.

\begin{figure}
  \begin{center}
    \includegraphics[width=0.5\textwidth]{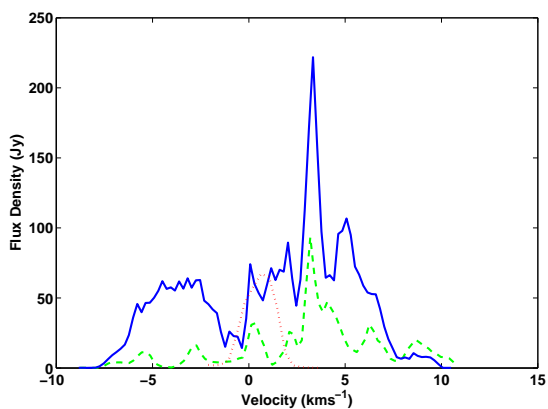}
  \end{center}
   \caption{Maser Spectra for \water\ (red dotted line) and SiO $v$=2
     (green dashed line) and $v$=1 (blue solid line) determined from
     the sum of the model components fitted to the image cube.}
\label{fig:K-maser-spec}
\end{figure}

With these calibration steps completed we (Column 2, Fig. \ref{fig:flow}) 
calibrated and separated the IF with the spectral line R\,LMi \water\ maser data and
applied the doppler corrections using the LSR standard of rest and
Radio velocity definitions (with SPLAT and CVEL) \citep{vlba_spec}. Additional
corrections to the amplitudes could be derived with ACFIT
at this point; these were the order of 10\% and are possibly from
the uncorrected atmospheric opacity. 
The 22GHz phase referenced spectral line data could then be imaged and
a channel with a compact, strong, point source dominated emission was
selected for further analysis. It was immediately obvious where the
peak of the brightness fell in the image and a single model component
at the site of this peak was used for a phase refinement of the data,
generated by CALIB, followed by re-imaging and deconvolution.
This latter step, which is mainly correcting for the differential
static atmospheric contributions, would loose the astrometrical
registration if we were not able to find a reasonable apriori position
for the conventionally phase referenced data and use that for the
calibration model. Therefore one of the requirements for successful
non-integer SFPR is that an adequate initial phase referenced image can be
formed. As the accuracy we require for this position is a few mas this
is not an overly demanding requirement.
The outcome is that the first step produces a conventional PR image of the
\water\ masers, and the second provides phase refinements to that
dataset whilst retaining the location of the maximum
emission. Therefore the \water\ maser cube is registered to the 4C39.25
calibrator. 
Figure \ref{fig:K-maser-spec} plots the spectra of the recovered
emission (i.e. the flux density in the clean components) from the
\water\ image cube with a red dotted line.

The 43GHz 4C39.25 data (Column 3, Fig. \ref{fig:flow}) was first adjusted so that the
reference frequency value was set, using AIPS task PUTHEAD, to 43.620\,GHz; that is double the
reference frequency value of the lower frequency and outside the
observed band. The frequency axis reference pixel
was also changed. This ensured the labelling of the data was correct
while preserving an integer reference frequency ratio and the phases could be doubled without
considerations of the ambiguities. These two steps are specifically 
required for non-integer frequency ratio analysis. Then we doubled the
value of phases measured from the delay observations of 4C39.25 at 22\,GHz, using our own external
script, to produce a FPT dataset. As a word of warning to users we
mention that we do not recommend the AIPS 
task SNCOR for this purpose, as it does not do a good job when there are
multiple IFs unless the separations between the IFs also scale by the
same frequency ratio factor. However, it functions correctly when there is only
one IF per frequency band. We calibrated the remaining dispersive phase
corrections, which were changing on $\sim$hour long timescales, with
CALIB. 

\begin{figure}
  \begin{center}
    \includegraphics[width=0.9\textwidth]{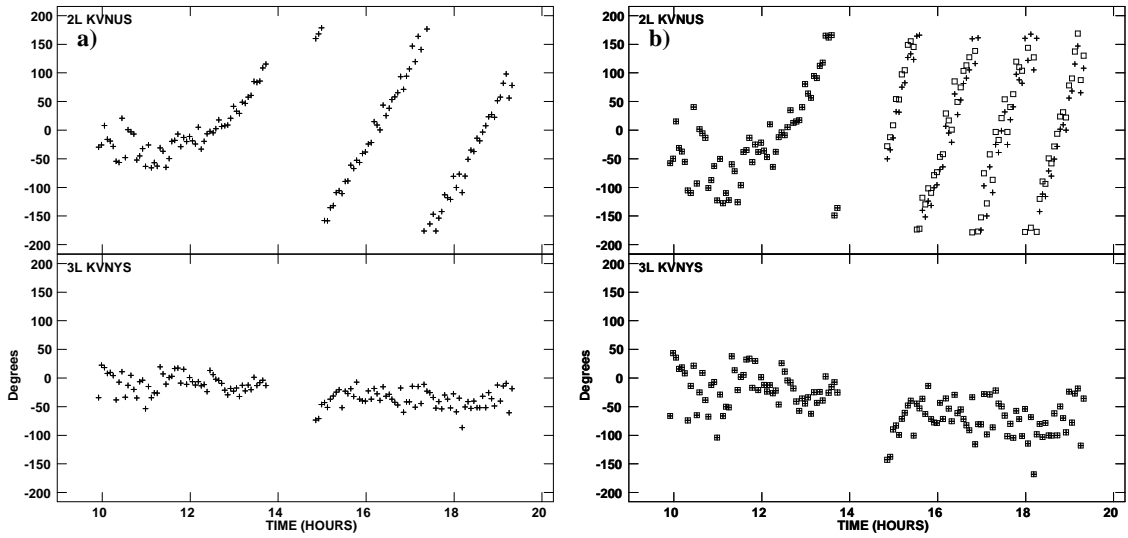}
  \end{center}
  \caption{Plots of a) the antenna-based phase-residual corrections to
    the FPT data, on the \water\ maser observed at 22GHz, derived by
    self-calibration with CALIB and b) the corrections derived for the
    SiO $v$=1 maser by scaling with the non-integer frequency ratio
    between the two transitions both before correcting for the dropped
    turn of phase at zenith (crosses) and after correcting for the
    dropped turn of phase (squares). The shift between the two
    solutions is 22$^o$.}
\label{fig:snplt}
\end{figure}

Once again (Column 4, Fig. \ref{fig:flow}) we calibrated and separated
those IFs with the R\,LMi
SiO maser data using SPLAT, using the scaled 4C39.25 delays from 22GHz and
the dispersive phase corrections from 4C39.25 at 43GHz.  The doppler and
amplitude corrections were calculated with CVEL and ACFIT as before.
We transferred the phase solutions from the \water\ maser and
scaled these values by the frequency ratio between the \water\ and the
SiO maser $v$=1 and $v$=2, that is 1.939 and 1.926 respectively. This
data was then ready for imaging. However the images were not of very
good quality and looking back at the solutions it was clear that in
the observing gap at zenith, a turn of phase had been missed on the
Ulsan antenna in the second half of the experiment.
We added in the phase jump that would have been introduced by this missed turn of phase
($2\pi\,(2-\nu_{\rm SiO}/\nu_{H_2O})$ for the two transitions, equal to 27$^o$
and 22$^o$ respectively) using SNCOR. Figure \ref{fig:snplt} shows the
measured phase corrections (i.e. the phase residuals) for the \water\  maser and the derived corrections for the
SiO $v$=1 maser before and after adding in the lost phase turn.
With these steps completed we were able to directly image the SiO
masers at the two transitions, which were now accurately registered to the \water\  emission
and that in turn to 4C39.25. We imaged the data via the normal
methods. Given the limited number of baselines we found the most
reliable deconvolution came with fitting a small number of model
components to the data, which we did in difmap \citep{difmap}. 
Figure \ref{fig:K-maser-spec} plots with a blue and green solid and dashed lines the spectra of the recovered
emission (i.e. the flux density in the clean components) from the SiO image cubes.
The images are shown in Section \ref{sec:res}. Plotting of the cubes
for the figures in this paper was done with Miriad \citep{miriad}.

\section{Results}
\label{sec:res}

\subsection{Astrometric Imaging}

Our analysis followed the steps required to produce PR maps of the \water\
maser emission, with respect to 4C39.25 and SFPR maps of the two
transitions of SiO maser
emission, with respect to the \water\ maser emission.
We present  images of the astrometrically registered velocity averaged
(moment zero) cubes of the \water\ and SiO maser emission  (Figure
\ref{fig:mom0}) and astrometrically registered velocity cubes of the two
transitions of  SiO maser emission  (Figure
\ref{fig:chan}). The image coordinates are relative to the peak of the SiO
maser emission in $v$=1. 
On Figure \ref{fig:mom0} we have overlaid the position of the optical
source R\,LMi, as measured by Hipparcos, for the observing epoch
(2011.17) with 1$\sigma$ errors (shown with a
large cross) \citep{hip07}. 
This position is listed in Table \ref{tab:pos}, which summarises all
the astrometric positions we have measured. The Hipparcos position has
errors that are dominated by the measurement errors in the proper
motion, multiplied by the two decades since those observations were
made. 
%
%
We measured the peak of the \water\ emission with IMFIT in the moment
zero map, which is listed in Table \ref{tab:pos} with the formal errors
on that fit. The much larger systematic errors are discussed in the following
section. Similarly the peak of the SiO $v$=1 emission was measured and
the position and errors are in the same table. 
We fitted a ring to a mask of the combined $v$=1 and 2 emission, by
a least-square maximisation of the overlap of a ring model with the
pixels in the image in either transition with emission greater than
10\% of the peak in that transition. 
%
%
The centre of this ring should be the site of the centre of optical emission of
R\,LMi. We take the width of the ring as the uncertainty in that position. The ring
is shown as the circle, with the centre and the uncertainty shown with
a small cross, in both Figures \ref{fig:mom0} and \ref{fig:chan}. The values are listed in Table \ref{tab:pos}.  

\begin{figure}
  \begin{center}
    \includegraphics[width=0.5\textwidth,angle=-90]{fig5.eps}
  \end{center}
  \caption{Contour plots of the Moment Zero integrated SFPR maps with
    absolute astrometry relative to the peak of the SiO
    $v$=1, J=0$\rightarrow$1 emission centred at
    09:45:34.284 +34:30:42.765 (epoch 2011.7 in equinox J2000 coordinates). The $v$=1  SiO
    emission (in blue) has contour levels of 10 to 80
    Jy/bm\,\kms\ in steps of 10, and the $v$=2 (green) has contour
    levels of 3 to 24 Jy/bm\,\kms\ doubling at every step. The
    \water\ maser emission (red) is overlaid with contour levels of 3, 6,
    12, 16 20 \& 24 Jy/bm\,\kms . The boundary and the centre of the ring fitted to the SiO
    emission, and the uncertainties in the fitting, is included as the circle
    with small cross at centre. This would be expected to centre on the
    R\,Leo Minoris optical emission. The large
    cross indicates the Hipparcos
    position for this source, with the bar length being the proper
    motion error over the 21 years since the optical positional
    epoch. The beam for the two bands appears in the bottom left.}
\label{fig:mom0}
\end{figure}

\begin{figure}
  \begin{center}
    \includegraphics[width=0.95\textwidth,angle=-90]{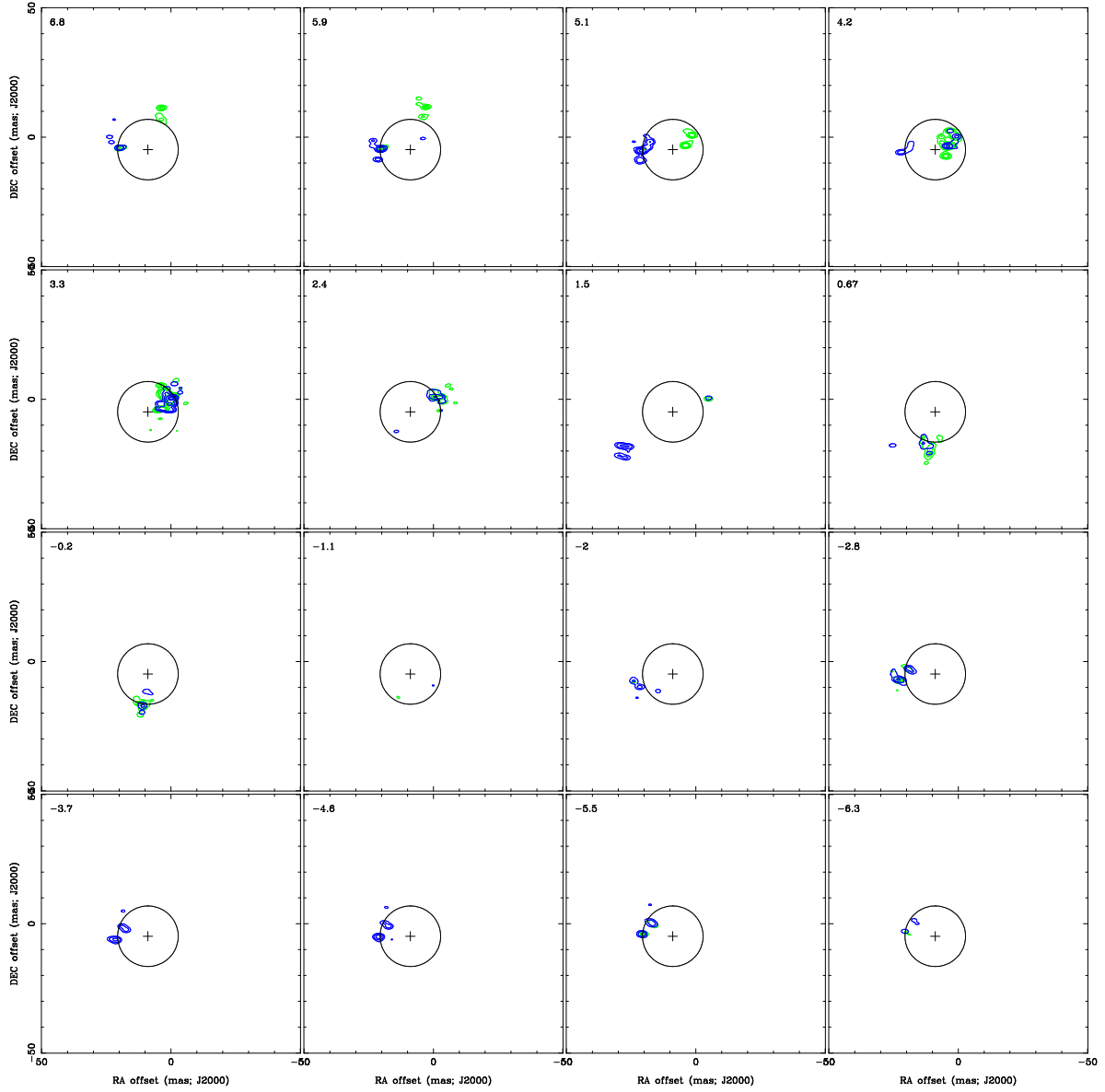}
  \end{center}
   \caption{Contour plot of the Velocity Channels map of the SiO emission in
     $v$=1 (blue) and $v$=2 (green) for 
     J=0$\rightarrow$1 centred on the peak emission. The 
     circle for the SiO emission from Figure \ref{fig:mom0} is
     included. The data are averaged to 0.9\kms\ channels with
     contours at 1, 2, 3, 6, 12, 24 and 32 Jy\,\kms\ for $v$=2 and at 3,
     6, 12, 24 and 50 Jy\,kms\ for $v$=1. The average velocity is given
     in the top left of each panel.} 
\label{fig:chan}
\end{figure}

\begin{table}
\begin{tabular}{|l|llll|}
\multicolumn{5}{c}{Astrometric Positions of Sources at Epoch 2011.17}\\
\hline
& $\alpha_{\rm 2000}$&$\Delta \alpha$&  $\delta_{\rm 2000}$&$\Delta \delta$\\
& h:m:s&mas&  d:m:s &mas\\
\hline
Predicted Hipparcos position of R\,LMi&09:45:34.2867& 42& +34:30:42.718 &25\\
Peak of \water\ maser emission&09:45:34.2890& 0.6 &+34:30:42.771 & 0.4\\
Peak of the SiO $v$=1 emission & 09:45:34.2838 & 0.4& +34:30:42.765 & 0.4 \\
Centre of SiO emission ($v$=2 and 1) & 09:45:34.2845&\phantom{0}5 & +34:30:42.759 &\phantom{0}5\\
\hline
\multicolumn{5}{c}{Implied Proper Motion of R\,LMi}\\
\hline
&$\mu_{\alpha {\rm cos}\delta} $&$\Delta \mu_{\alpha {\rm cos}\delta}$&$\mu_\delta$&$\Delta\mu_\delta$\\
& mas/yr&mas/yr&mas/yr &mas/yr\\
\hline
Combined Hipparcos and VLBI observations& 2.3 & 0.3 & -3.9 & 0.3\\
\hline

\end{tabular}
\caption{Astrometric Positions of R\,LMi for optical, \water\ and SiO
  emission for epoch 2011.17, with formal 1-$\sigma$ errors. Assuming the Hipparcos position of epoch
  1991.25 and the centroid of the SiO of epoch 2011.17 are compatible
  provides an improved proper motion value.}
\label{tab:pos}
\end{table}

%

\subsection{Error Analysis}


The errors in Table \ref{tab:pos} are the formal errors in the fitting
to the data. For the VLBI data these are completely
dominated by the systematic effects, which we review here.

The \water\ maser peak emission position is conventionally phase
referenced to 4C39.25, which has a very accurate absolute
reference position with errors of about 0.1mas \citep{4c39}, but an angular
separation of 5.9$^o$ from the R\,LMi target. 
Because of this large separation this would not normally be considered
a good phase referencing calibrator.  With such an observing setup,
even in good weather, one would expect of the order of 30$^o$ of phase
noise from the dynamic tropospheric contributions, more than 100$^o$
of phase noise from the static tropospheric contributions, 15$^o$ of
phase noise from the static ionospheric contributions and less than
1$^o$ of phase noise from the dynamic ionospheric contributions
(following the formulae in \citet{a07}). However we note that the
atmospheric coherence of the KVN appears to be much better than those
for the typical cases \citep{sslee_kvn}. This maybe due to better
coherence between the atmosphere at the different sites, which could
be from the short baselines or that the weather moves down off the
Chinese landmass with little disruption. 
Various methods exist to reduce the dominant static tropospheric
contributions, as reviewed in \citet{trop-correct}, but these were not
followed in this experimental setup as we did not originally intend to attempt to
achieve absolute astrometric connection. Nevertheless on inspection of
the data we realised that should be possible, as the peak of the brightness was clearly
identifiable.
To confirm that errors were minor we compared the peak flux before and
after self calibration. This gives a measure of the scale of the
random errors. If these errors are small one can be confident that the
data is largely coherent and the astrometry is preserved. The
fractional flux recovery we found for the whole data set was 86\%, and
96\% for the data taken in the first half of the experiment. 
For our analysis we took the postion of the water maser peak from a model fit of a
single component to the data, limited to regions where the phase
residuals were smallest. That is the data between Yonsei and Tamna,
plus Ulsan in the first half of the experiment, which had a fractional flux
recovery of 91\%.
Given the large angular separation between the sources we take a
generous upper bound for the positional errors to be the resolution,
5$\times$3 mas. In future work we will attempt to improve on these limits.

The optical star R\,LMi has a relatively poor a priori position from
Hipparcos.  The uncertainty is in the proper motion measurement which,
given that the Hipparcos data is referenced to the observational epoch
of 1991.25, results in a position error of $\sim$50\,mas. However as
we phase reference the \water\ maser emission to 4C39.25 the
positional errors in our analysis are less.
%
Taking this as the dominant source of error in our analysis we use it
to determine the final error bars in our astrometric measurements
between the maser transitions. We use the relationship of $\Delta
\nu/\nu . \Delta \theta$, where $\Delta \theta$ is the astrometric
accuracy of the reference (i.e. the \water\ emission) and $\Delta \nu$
is the frequency gap between the transitions. In this case the
registration would, with $\Delta \theta$ of 5\,mas, have an
uncertainly of 2.5\,mas between
the \water\ and the SiO and 35\,$\mu$as between the SiO $v$=1 and 2
masers. In both cases the typical separations between, firstly the
\water\ and the SiO masers and the second case between the SiO  $v$=1
and 2 masers, are much greater than these errors, which will allow us
to have faith in any separations detected. SiO rings are close to the
surface of the star and water masers maybe ten times further out  (10$^{15}$cm),
which implies a typical separation of 60 mas between the sites of the
\water\ and the SiO maser emission for a source at
1\,kpc; that is the astrometric errors would be the order of 4\%. An error in
astrometry of 35\,$\mu$as between the SiO maser transitions implies a
physical scale of 5$\times 10^{11}$cm at 1\,kpc, or 0.5\% of the typical ring
diameter (10$^{14}$cm).

The retention of absolute astrometric registration in the VLBI data
allows us to compare the centre of the SiO maser emission, where the
star in R\,LMi should be, and the Hipparcos position for this
star. Measuring from the astrometric image derived, we
find the position offset between these to be -33 and 41 mas. 
The error budget on this measurement consists of the errors in 
the fit to the centre of the circle, which is 5\,mas, combined
with the absolute astrometric accuracy of the \water\ emission gives a
total estimated error of 7\,mas.
Given the twenty years since the Hipparcos reference epoch the errors
in proper motion from the Hipparcos data dominate. The centre of the
circle of SiO maser emission falls within the 1$\sigma$ error
($\sigma_{\rm RA_{\rm pm}}$=35 mas) in the Hipparcos
value of along $\mu_{\alpha {\rm cos}\delta}$ and within the
2$\sigma$ error ($\sigma_{\delta_{\rm pm}}$=25 mas) in the Hipparcos value of along $\mu_\delta$
\citep{hip07}.
Forthcoming observations from the GAIA satellite \citep{gaia} will be
able to confirm this VLBI result.

Combining the VLBI position we have derived in 2011.17, with
uncertainties in the fit and those of the \water\ maser emission to
give a total of 7\,mas, with that of Hipparcos from 1991.25, with
uncertainties of 1\,mas, we can derive the proper motion of R\,LMi
over the twenty year baseline and this is given in Table
\ref{tab:pos}.

Subtracted from (both) the reported VLBI positions would be the core
shift (and any other structure phases) from 4C39.25 which, because it
was assumed to be achromatic in the analysis, would contaminate these
results. These effects are far below the levels of accuracy achieved
in this analysis but will become significant when global baselines are
included.

\section{Discussion}
\label{sec:dis}

We report here astrometrically aligned images of the \water\
and SiO masers around R\,LMi. This is the first demonstration of the SFPR method with
non-integer frequency phase ratios, and the first demonstration of the
combination of SFPR with PR to retain absolute astrometry through out the
analytical chain. Our error analysis cautiously sets large errors on
the absolute position of the \water\ maser emission, and these errors
are matched by the uncertainty in the ring fitting to find the centre of
the SiO maser emission, but these are still a major improvement in the
uncertainty in the Hipparcos position for this source.

There have been previous attempts to make astrometrical observations
for AGB stars with SiO and \water\ masers. A limited number were made
with VERA and so provide bona-fide astrometry, the others do
not. Now, with the new methods discussed here, mm-VLBI astrometric line
experiments will become much more straight forward and, we hope,
widely implemented.
We will, now the method is understood, undertake a major monitoring
campaign of interesting AGBs hosting \water\ and SiO masers, and
massive star forming regions hosting \water\  and \methanol\ masers --
potentially up to and including the 95GHz line. To aid other users who
wish to follow the same analysis we provide a list of what we
consider the required conditions for a successful outcome.

\subsection{Crucial considerations for success}
\label{sec:suc}

\begin{itemize}
\item SFPR requires that  the calibrator is within the same
  ionospheric region
  (the `patch') for dispersive contributions. Fortunately the patch size scales
  with frequency so for high frequencies this will be very large. The
  actual scale size at mm-frequencies is under investigation;
  extrapolating from measurements at lower frequencies (e.g. \citet{tms}) is not
  possible as the approximations are no longer valid. In \citet{rioja_11a} we have
  successfully used calibrators 10$^o$ from the target.
\item SFPR for spectral-line sources must use the delays derived from
  the continuum calibrator source,
  as the line of sight corrections to the delays are not derivable on
  the line source. Therefore the antenna position errors must be
  sufficiently small (less than a few cm) so that they do not introduce rapid phase changes which could
  not be tracked on the other source. %
\item Non-integer frequency ratio SFPR requires the delay solutions to
  be of sufficient quality to allow extrapolation to a reference point
  outside of the observing band. A 0.05 nsec error in the delay would
  introduce a 9$^o$ error in the phase for a point 500MHz from the
  frequency reference pixel. We suggest that one should try and ensure
  this extrapolation is no further than the bandwidth spanned.
\item Non-integer frequency ratio SFPR requires that the phase solutions can be tracked
  across wraps of phase ambiguities, so continuous observations
  without gaps are desirable.  
\item Non-integer frequency ratio SFPR also requires that the
  positions of the target source at the lower frequency are
  accurate. The absolute error in the target position at the reference
  frequency, when applied to the target frequency, is diluted by the
  fractional frequency span $\Delta \nu/\nu$. For SFPR this fractional
  span approaches integer so that the errors in the registration
  between the frequencies can approach the absolute positional
  error. If needed the positions can be derived by conventionally
  phase referencing the low frequency data. In this case the source
  switching needs to be sufficiently fast to allow phase referencing
  at the low frequency.
\item For arrays which don't support (yet) simultaneous multichannel
  receivers (e.g. the VLBA) fast frequency switching will still work
  with this approach. Fast frequency switching is limited in the
  maximum frequencies which can be supported (as the variation in the rates can not be
  higher than the cycle time of the frequency switching). In our
  experience fast frequency switching between frequencies as high as 43 and 86\,GHz is
  achievable in good conditions with switching cycle period of 1
  minute. This is about the maximum speed possible with the VLBA
  sub-reflector rotator.

\end{itemize}

\section{Conclusions}
\label{sec:con}

We have demonstrated that the KVN is capable of performing non-integer
Source Frequency Phase Referencing on simultaneously observed line
sources, between \water\ and SiO in this particular case. The method
would also be suitable for \water\ and 44GHz \methanol\ targets. This
would provide astrometric positions between all the transitions, and
potentially (if the \water\ masers are phase referenced) absolute
astrometric positions for all the transitions, as in this case.

Our method allows a high quality mas-level astrometric alignment of the two SiO 
frequencies with respect to the \water\  emission. The registration to
4C39.25 and the absolute coordinate system is based on how far one
trusts the conventional phase referencing. We have explained why we
believe that the errors in the \water\ maser position are less than 5\,mas.
This provides a much improved position for the centre of a ring containing the SiO
emission, which would be the site of the optical source.  This position
is within the errors from the Hipparcos observations. The error on the fit to
the centre of the circle is 5\,mas, which when combined with the
absolute astrometric accuracy of the \water\ emission gives a total
estimated error of 7\,mas.
From the new position we are able to improve on the proper motion
derived by Hipparcos for the optical star R\,LMi, improving the
precision by nearly an order of magnitude, to 0.3\,mas/yr.

The registration between the two SiO transitions is expected to be the
order of 35\,$\mu$as, which will allow for very exact separation of the
location of the different masing components. The registration of the
\water\ and the SiO maser emission has an error of 2.5\,mas.
Detailed analysis of the SiO emission structure is left for future
publications, when we have monitored the SiO masers through the full
cycle of the pulsation period. Additionally in the near future we are
hopeful of being able to access longer baselines to Japan, Spain or
Australia, which would provide a desirable boost in resolution.

\bigskip
\noindent
{\bf Acknowledgements}

\noindent
We are grateful to all staff members and students in the KVN who
helped to operate the array.  The KVN is a facility operated by the
Korea Astronomy and Space Science Institute.  RD acknowledges the
support of the Korean Ministry of Science, ICT \& Future Planning
Brainpool Fellowship (121S-1-2-0228).
We acknowledge with thanks the variable star observations from the
AAVSO International Database, contributed by observers worldwide, that
provided the optical phase for our observations.
We wish to thank the anonymous referee, whose comments improved
the presentation of the material in this paper.  



\begin{thebibliography}{}


\bibitem[Alef(1988)]{alef} Alef, W.\ 1988, The Impact  of  VLBI on Astrophysics and Geophysics, 129, 523 


\bibitem[Asaki et al.(2007)]{a07} Asaki, Y., Sudou, H., 
Kono, Y., et al.\ 2007, \pasj, 59, 397 


\bibitem[Beasley \& Conway(1995)]{vlba_pr} Beasley, A.~J., \& Conway, J.~E.\ 1995, Very Long Baseline Interferometry and the VLBA, 82, 327 

\bibitem[Benson \& Little-Marenin(1996)]{water_agb} Benson, P.~J., \& Little-Marenin, I.~R.\ 1996, \apjs, 106, 579 

\bibitem[Boboltz \& Wittkowski (2005)]{bob_05} 
Boboltz, D.~A., \& Wittkowski, M.\ 2005, \apj, 618, 953 

\bibitem[Bowen(1988)]{bowen88} Bowen, G.~H.\ 1988, \apj, 329, 299 

\bibitem[Bujarrabal (1994)]{buj94} Bujarrabal, V.\ 1994, \aap, 285, 971 






\bibitem[Choi \etal (2008)]{choi_08} Choi Y. K., Hirota T., Honma
  M., Kobayashi H., Proceedings of the 9th European VLBI Network
  Symposium, 
  ``Distance to VY Canis Majoris with VERA''. SISSA, Trieste;
  2008. 

\bibitem[Cooke \& Elitzur(1985)]{elitzur_water} Cooke, B., \& Elitzur, M.\ 1985, \apj, 295, 175 

\bibitem[Cotton et al.(2011)]{cot_11} Cotton, W.~D., Ragland, 
S., \& Danchi, W.~C.\ 2011, \apj, 736, 96 

\bibitem[Cotton et al.(2010a)]{cot_10a} Cotton, W.~D., Ragland, 
S., Pluzhnik, E.~A., et al.\ 2010, \apjs, 188, 506 

\bibitem[Cotton et al.(2010b)]{cot_10b} Cotton, W.~D., Ragland, 
S., Pluzhnik, E.~A., et al.\ 2010, \apjs, 187, 107 

\bibitem[Cotton et al.(2009a)]{cot_09a} Cotton, W.~D., Ragland, 
S., Pluzhnik, E.~A., et al.\ 2009, \apjs, 185, 574 

\bibitem[Cotton et al.(2009b)]{cot_09b} Cotton, W.~D., Ragland, 
S., Pluzhnik, E., et al.\ 2009, \apj, 704, 170 

\bibitem[Cotton \etal (2008)]{cot_08} Cotton, W.~D., Perrin, G., \& 
Lopez, B.\ 2008, \aap, 477, 853 

\bibitem[Cotton \etal (2006)]{cot_06} Cotton, W.~D., et al.\ 2006, 
\aap, 456, 339 

\bibitem[Cotton \etal (2004)]{cot_04} Cotton, W.~D., et al.\ 2004, 
\aap, 414, 275 


\bibitem[Cotton (1995)]{cotton_fring_95} Cotton, W.~D.\ 1995, Very Long 
Baseline Interferometry and the VLBA, 82, 189 

\bibitem[Desmurs \etal\ (2000)]{desmurs00} Desmurs, J.~F., 
Bujarrabal, V., Colomer, F., \& Alcolea, J.\ 2000, \aap, 360, 189 

\bibitem[Deller et al.(2011)]{difx} Deller, A.~T., Brisken, 
W.~F., Phillips, C.~J., et al.\ 2011, \pasp, 123, 275 

\bibitem[Dodson \& Rioja(2009)]{vlba_31} Dodson, R., \& Rioja, M.~J.\ 2009, VLBA
Science Memo \#31 

\bibitem[Dodson \& Rioja (2011)]{dodson_eavw_11} Dodson, R., Rioja, M.~J., Jung T.H., 
2011, East Asia VLBI Workshop, Lijiang, China. 

\bibitem[Elitzur(1980)]{elitzur_80} Elitzur, M.\ 1980, \apj, 240, 553 

\bibitem[Elitzur(1992)]{elitzur_92} Elitzur, M.\ 1992, \araa, 30, 75 


\bibitem[Fey et al. (2004)]{4c39}Fey, A.L., et al., 2004, \aj, 127, 3587

\bibitem[Fomalont (2014)]{alma_test}Fomalont, E., 2014, Personal Comms.


\bibitem[Han et al.(2008)]{han_08} Han, S.-T., Lee, J.-W., 
Kang, J., et al.\ 2008, International Journal of Infrared and Millimeter 
Waves, 29, 69 

\bibitem[Han et al.(2013)]{kvn_optics} Han, S.-T., Lee, J.-W., 
Kang, J., et al.\ 2013, \pasp, 125, 539 

\bibitem[Honma \etal\ (2008)]{trop-correct} Honma, M., Tamura, Y., 
\& Reid, M.~J.\ 2008, \pasj, 60, 951 


\bibitem[Humphreys \etal\ (2000)]{hum2000} Humphreys,
E.~M.~L., Gray, M.~D., Yates, J.~A., Field, D., Bowen, G., \& Diamond,
P.~J.\ 2000, EVN Symposium 2000, Proceedings of the 5th european VLBI
Network Symposium, 197

\bibitem[Humphreys(2002)]{hump02} Humphreys, E.M.L., Gray, M.D., Yates,  J.A., \etal\ 2002, \aap, 385, 256 

\bibitem[Gray et al.(1999)]{gray_99} Gray, M.~D., Humphreys, 
E.~M.~L., \& Yates, J.~A.\ 1999, \mnras, 304, 906 

\bibitem[Gray et al.(2009)]{gray_09} Gray, M.~D., Wittkowski, M., Scholz, M., et al.\ 2009, \mnras, 394, 51 


\bibitem[Griesen (2003)]{aips} Griesen, E. W., 2003, in Information 
Handling in Astronomy -- Historical Vistas, Heck, A. ed., Kluwer 
Academic Publishers, Dordrecht, ISBN 1-4040-1178-4, Astrophysics and
Space Science Library, 285, 109.


\bibitem[Lee et al.(2014)]{sslee_kvn} Lee, S.-S.,  et al.\ 2014, \aj, 147, 77


\bibitem[Kamohara et al. (2010)]{kamohara_10} Kamohara, R., Bujarrabal, V., Honma, M., et al.\ 2010, \aap, 510, A69 

\bibitem[Keenan et al. (1974)]{keenan_74} Keenan P.C., Garrison, R.F.,
Deutsch, A.J., 1974, \apjs, 28, 271

\bibitem[Kim et al.(2010)]{sio_h2o_kasi} Kim, J., Cho, S.-H., Oh, C.~S., \& Byun, D.-Y.\ 2010, \apjs, 188, 209 


\bibitem[Oh et al. (2011)]{kvn_backend} Oh, S.-J., \etal , 
2011, Publications of the Astronomical Society of Japan, 63, 1229



\bibitem[Pardo \etal\ (2004)]{pardo_04} Pardo, J.~R., Alcolea, J.,
Bujarrabal, V., Colomer, F., del Romero, A., \& de Vicente, P.\ 2004,
\aap, 424, 145

\bibitem[Perryman \etal\ (1997))]{perry_97} Perryman, M.A.C., et  al. 1997, \aap, 323, L49

\bibitem[Perryman \etal\ (2001)]{gaia} Perryman, M.~A.~C., de Boer, K.~S., Gilmore, G., et al.\ 2001, \aap, 369, 339 

\bibitem[Pickle \& Depagne (2010)]{RLMi_distance}{{Pickles}, A. \& {Depagne}, {\'E}.}, 2010, \pasp, 122, 1437

\bibitem[Reid (1995)]{vlba_spec} Reid, M.~J.\ 1995, Very Long Baseline Interferometry and the VLBA, 82, 209 


\bibitem[Richter et al.(2013)]{richter13} Richter, L., Kemball, A., \& Jonas, J.\ 2013, \mnras, 436, 1708 


\bibitem[Rioja et al.(2011)]{rioja_11b} Rioja, M., Dodson, R., 
Malarecki, J., \& Asaki, Y.\ 2011, \aj, 142, 157 

\bibitem[Rioja \& Dodson(2011)]{rioja_11a} Rioja, M., \& Dodson, R.\ 2011, \aj, 141, 114 

\bibitem[Rioja \& Dodson(2009)]{vlba_32} Rioja, M.~J., \& Dodson, R.\ 2009, VLBA
Science Memo \#32 


\bibitem[Rioja et al.(2008)]{rioja_08} Rioja, M.~J., Dodson, R., 
Kamohara, R., et al.\ 2008, \pasj, 60, 1031 

\bibitem[Rioja et al.(2014)]{rioja_14} Rioja, M.~J., Dodson, R., et
  al.\ 2014, Manuscript AJ-12041

\bibitem[Schwab \& Cotton(1983)]{cotton_schwab} Schwab, F.~R., \& Cotton, W.~D.\ 1983, \aj, 88, 688 

\bibitem[Shepherd et al.(1994)]{difmap} Shepherd, M.~C., 
Pearson, T.~J., \& Taylor, G.~B.\ 1994, \baas, 26, 987 

\bibitem[Soria-Ruiz \etal\ (2004)]{soria04} Soria-Ruiz, R., 
Alcolea, J., Colomer, F., Bujarrabal, V., Desmurs, J.-F., Marvel, K.~B., \& 
Diamond, P.~J.\ 2004, \aap, 426, 131 

\bibitem[Soria-Ruiz \etal\ (2005)]{soria05} Soria-Ruiz, R., 
Colomer, F., Alcolea, J., Bujarrabal, V., Desmurs, J.-F., \& Marvel, K.~B.\ 
2005, \aap, 432, L39 

\bibitem[Soria-Ruiz \etal\ (2007)]{soria07} Soria-Ruiz, R., 
Alcolea, J., Colomer, F., Bujarrabal, V., \& Desmurs, J.-F.\ 2007, \aap, 
468, L1 

\bibitem[Sault et al.(1995)]{miriad} Sault, R.~J.,
  Teuben,  P.~J., \& Wright, M.~C.~H.\ 1995, Astronomical Data Analysis Software and Systems IV, 77, 433 

\bibitem[Thompson et al.(2001)]{tms} Thompson, A.~R., 
Moran, J.~M., 
\& Swenson, G.~W., Jr.\ 2001, ''Interferometry and synthesis in radio astronomy by A.~Richard Thompson, James M.~Moran, and George W.~Swenson, Jr.~2nd ed.~ New York : Wiley

\bibitem[van Leeuwen(2007)]{hip07} van Leeuwen, F.\ 2007, 
Astrophysics and Space Science Library, 350,  


\bibitem[Whitelock \etal\ (2008)]{PL-rel} Whitelock, P., Feast,
  M., \& {van Leeuwen}, F. \ 2008, \mnras, 386, 313

\bibitem[Yi \etal\ (2005)]{yi05}
  Yi, J., Booth, R.S., Conway, J.E., Diamond, P.J., 2005, A\&A, 432, 531



\end{thebibliography}
\end{document}